\documentclass[aps, noshowpacs,preprintnumbers,amsmath,amssymb]{revtex4}

\usepackage{verbatim}
\usepackage{wasysym}

\usepackage{graphicx}
\usepackage{epstopdf}
\usepackage{dcolumn}
\usepackage{bm}
\usepackage{wrapfig}

\begin{document}

\title{Metal nanoparticle field-effect transistor}

\author{Yuxue Cai}
\affiliation{Institute of Physical Chemistry, University of Hamburg, 20146 Hamburg, Germany}

\author{Jan Michels}
\affiliation{Physics and Chemistry Departments, Interdisciplinary Nanoscience Center Hamburg, University of Hamburg, Sedanstrasse 19, 20146 Hamburg, Germany}

\author{Julien Bachmann}
\affiliation{Physics and Chemistry Departments, Interdisciplinary Nanoscience Center Hamburg, University of Hamburg, Sedanstrasse 19, 20146 Hamburg, Germany}
\affiliation{Chemistry Department, University of Erlangen-Nurnberg, Egerlandstrasse 3, 91058 Erlangen, Germany}

\author{Christian Klinke}
\email{klinke@chemie.uni-hamburg.de}
\affiliation{Institute of Physical Chemistry, University of Hamburg, 20146 Hamburg, Germany }

\begin{abstract} 

We demonstrate that by means of a local top-gate current oscillations can be observed in extended, monolayered films assembled from monodisperse metal nanocrystals -- realizing transistor function. The oscillations in this metal-based system are due to the occurrence of a Coulomb energy gap in the nanocrystals which is tunable via the nanocrystal size. The nanocrystal assembly by the Langmuir-Blodgett method yields homogeneous monolayered films over vast areas. The dielectric oxide layer protects the metal nanocrystal field-effect transistors from oxidation and leads to stable function for months. The transistor function can be reached due to the high monodispersity of the nanocrystals and the high super-crystallinity of the assembled films. Due to the fact that the film consists of only one monolayer of nanocrystals and all nanocrystals are simultaneously in the state of Coulomb blockade the energy levels can be influenced efficiently (limited screening). \\


\end{abstract}

\maketitle

\section*{Introduction}

Isolated conductive particles with dimensions in the nanometer range possess very small (self-) capacities. Charging such islands with electrons requires a certain energy, due to Coulomb repulsion, which amounts to $E_{C} = e^{2}/2C$, with $e$ the elementary charge and $C$ the capacity of the island. At low temperatures the thermal energy of the electrons is not sufficient to overcome the charging energy. This leads to a virtually insulating character of the structures. The phenomenon is known as \textit{Coulomb blockade}. The blockade regime can be overcome either by applying a large enough bias at which electrons start to tunnel via the nanocrystal, or by thermally activated electron hopping. For a lithographically defined tunnel-junction this was demonstrated in the late 1980s by Fulton and Dolan \cite{FULTON1987}. Later, this has also been shown for one individual nanocrystal \cite{ANDRES1997} or a few ones \cite{KLEIN1996}. The introduction of a gate electrode leads to three-terminal devices which act as \textit{single electron transistors}. The gate voltage manipulates the energy level of the island. By tuning the gate voltage the number of electrons on the island can be adjusted. This was also demonstrated for one \cite{BEZRYADIN1997,WOLF2010} or a few \cite{SATO1997,BOLOTIN2004} individual nanocrystals. An overview about the foundations, prospects, and applications in the field of single-electron devices can be found in the review by Likharev \cite{LIKHAREV1999}.

In a macroscopic crystal with atoms as constituents the electronic behavior is mainly determined by the energy levels of the atoms, the coupling between adjacent sites, and the symmetry of the solid. These properties are not free to choose but given by nature. In assemblies of "artificial atoms" like nanocrystals it is possible to control these properties by the used material (work function), by the size and shape of the nanocrystals, by the coupling determined by the organic ligands, and by their assembly. The tunable character of the superstructures makes them interesting model systems for charge transport studies in confined systems \cite{TALAPIN2005,LIAO2006,MURRAY2007,TALAPIN2009} and a vast number of experimental and theoretical results have been published dealing with the transport through nanocrystal arrays \cite{STAVEREN1991,PENG1999,BLACK2000,BEVERLY2002,BEECHER2004,ANDRES2006}. For example, a decrease of the distance between adjacent nanocrystals increases the coupling between them and yields in a decrease of the Coulomb charging energy of the individual conductive islands. Eventually, this leads to a transition from insulating to metallic behavior (Mott-Hubbard transition) \cite{SIMON1998,MARKIVICH1999,QUINN2005}. Researchers have observed that in such films charging effects occur attributed to Coulomb blockade \cite{GORTER1951,GIAEVER1968,LAMBE1969}. The transport has been described by various theories \cite{NEUGEBAUER1962,ABELES1975,INGOLD1992}. Often the current-voltage curves of granular films can be fitted with the equation of Middleton and Wingreen \cite{MIDDLETON1993} $I(V) = A ((V-V_{T})/V_{T})^{\zeta}$. The threshold voltage $V_{T}$ and the exponent $\zeta$ depend on the dimensionality of the film. The non-linear increase in conductivity above the threshold voltage $V > V_{T}$ is attributed to the opening of conduction paths in the film. For roughly two-dimensional films the value for $\zeta$ has been found experimentally to be about 2.25 and for fractal films to be around 4.12 \cite{SUVAKOV2010,PARTHASARATHY2001,PARTHASARATHY2004}. In theoretical considerations values of $\zeta = 1$ have been predicted for one-dimensional structures such as chains of nanocrystals and a value of $\zeta = 5/3$ for two-dimensional films \cite{MIDDLETON1993}. 

Devices founded on new principles with promising new functions, higher performance, or reduced production costs should anyhow be reliable, large-scale producible, and CMOS compatible \cite{TAO2008}. One branch of material synthesis which fulfills these requirements is colloidal chemistry \cite{WELLER2003}. Today, the control over nucleation and growth phases of nanocrystals allows tayloring the crystal size, while the amount and nature of ligand molecules and precursors permits control over shape, surface properties, and atomic composition of the nanocrystals \cite{AHRENSTORF2007}. In some systems it is possible to synthesize such structures with almost atomic precision and in macroscopic amounts \cite{PARK2007,SHEVCHENKO2003}. The colloidal synthesis of nanocrystals is comparatively fast, inexpensive, and scalable. This enables one to spread nanocrystals easily and homogeneously by e.g. spin-coating onto flat surfaces such as silicon or silicon oxide \cite{MALYNYCH2002}. One method which is able to generate monolayered highly ordered films over vast areas is the Langmuir-Blodgett (LB) method \cite{ALEKSANDROVIC2008}. This delivers an experimental access to the percolative transport through such films \cite{ARAI2005,HOSOKI2008}. In a new approach we deposited nanocrystals onto diethylene glycol as subphase with an intermediate polarizability \cite{ALEKSANDROVIC2008,CAI2010,GRESHNYKH2008}. This methodology yields highly ordered metal nanocrystal films over unprecedented vast areas in the micrometer up to the centimeter range. 

We demonstrate that in monolayered films of well-organized metal nanocrystals current oscillations can be observed using a local top-gate realizing transistor function. Instead of the semiconductor band gap, the physical basis of the device is the Coulomb energy gap, a property which is tunable by the nanocyrstal size. We attribute the success to the high quality of the nanocrystals, of the film, and the atomic layer deposition (ALD) oxide that we used as gate dielectric. Since the film consists of only one monolayer of nanocrystals and all are simultaneously in the state of Coulomb blockade screening is limited and the energy levels can be influenced efficiently.

\section*{Experimental Section}

Cobalt-platinum nanocrystals were were synthesized by simultaneous reduction of platinum acetylacetonate (Pt(acac)$_{2}$) and thermal decomposition of cobalt carbonyl (Co$_{2}$(CO)$_{8}$) in the presence of 1-adamantane carboxylic acid (ACA) and hexadecylamine (HDA) as stabilizing agents. In a standard procedure, 65.6 mg Pt(acac)$_{2}$ was dissolved in 8 g HDA, 0.496 g ACA, 0.26 g 1,2-hexadecanediol (HDD) and 4 mL diphenylether at 65 $^{\circ}$C. When the solution turned clear, the mixture was degassed three times and heated to the injection temperature (165 $^{\circ}$C for the spherical and 155 $^{\circ}$C for the cubic nanocrystals). 92 mg Co$_{2}$(CO)$_{8}$ dissolved in 1.6 mL 1,2-dichlorobenzene were separately degassed three times at room temperature and injected quickly into the mixture under vigorous stirring. After stirring for one hour at the injection temperature, the temperature was increased to 230 $^{\circ}$C for two additional hours. The obtained particles were precipitated with 2-propanol, centrifuged, and re-suspended in toluene.

For the monolayer preparation by the Langmuir-Blodgett method the nanocrystals were washed two more times, re-suspended in toluene and spread onto the diethylene glycol subphase. After the toluene was evaporated, the nanoparticles were compressed by the barrier with a speed of 2 mm/min to a target pressure of 10 mN/m. The film was held at the target pressure for about two hours, allowing the film to relax and to rearrange. 

The oxide layers were prepare by ALD from trimethylaluminum (TMA) and water in a self-made ALD reactor using nitrogen as carrier gas. TMA was kept at room temperature, water at 40 $^{\circ}$C, and the samples were heated to 100 $^{\circ}$C. Each ALD cycle consisted of the following sequence: TMA pulse (0.1 s), exposure (20 s), purge (40 s), then water pulse (0.5 s), exposure (20 s), and purge (40 s). In these conditions, 546 ALD cycles correspond to approximately 77 nm of Al$_{2}$O$_{3}$, respectively. The thickness of the ALD layers was measured with a variable-angle spectroscopic ellipsometer by Dr. Riss Ellipsometerbau GmbH.

\section*{Results and discussion}

\begin{figure}[ht]
  \centering
  \includegraphics*[width=0.75\textwidth]{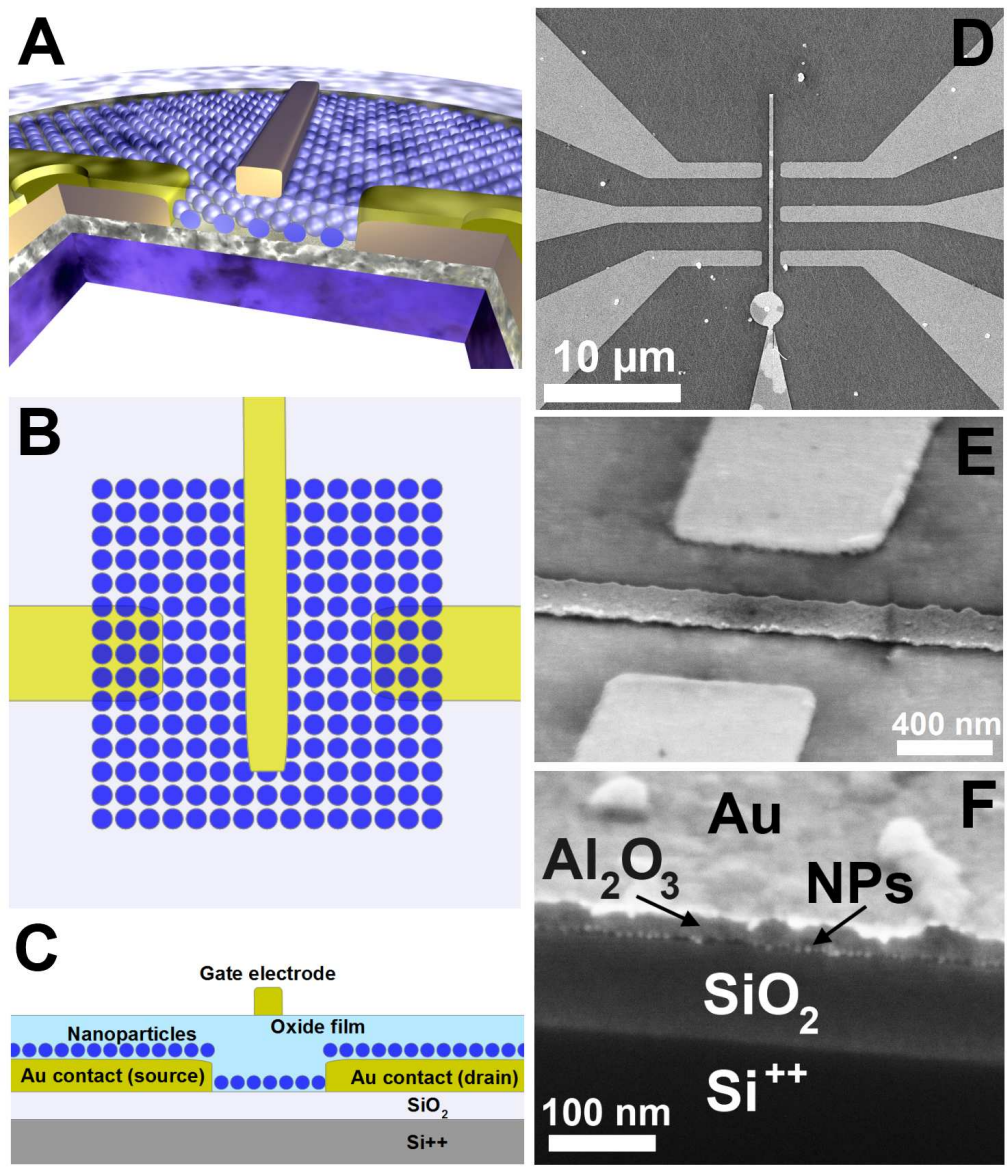}
  \caption{\textit{\textbf{Structure of the Coulomb blockade based transistor.} (A) 3D sketch of the device (Not to scale!); (B) Top view; (C) Cross-section sketch; (D) Top view SEM image: From left and right the gold electrodes are entering the view and defining three parallel devices. In the center the vertical structure is the top-gate electrode on the ALD Al$_{2}$O$_{3}$ film (perpendicular to the charge flow) which has no contact to the nanocrystal film; (E) Closed, tilted view: The top-gate electrode is clearly visible. The contacts are slightly blurred since they are located below the ALD film which is semi-transparent to the electron-beam; (F) Cross-sectional SEM image at the top-gate position showing all important components of the device.}}
\end{figure}

For the nanocrystals we chose a cobalt-platinum synthesis under Schlenk conditions which was published by Shevchenko et al. \cite{SHEVCHENKO2002} and applied in previous experiments \cite{ALEKSANDROVIC2008,CAI2010,GRESHNYKH2008}. It produces monodisperse nanocrystals with high chemical stability. We synthesized nanocrystals in two different sizes and shapes. The spherical nanocrystals have a diameter of 7.6 $\pm$ 0.5 nm and the cubic nanocrystals have 9.3 $\pm$ 0.8 nm. The particles were analyzed by energy dispersive X-ray analysis (EDX) to have a composition of Co$_{0.20}$Pt$_{0.80}$ (for further details see experimental section).

Figure S1 (see Supporting Information) shows a zoom series of TEM pictures of monodisperse spherical nanocrystals. The cobalt-platinum nanocrystals formed highly ordered monolayer films, wherein the particles were arranged in hexagonal domains. The quality of the nanocrystals has also been checked by X-ray diffraction (XRD). Figure S2 (see Supporting Information) shows a diffractogram of the spherical and the cubic nanocrystals. There are almost no differences between the two diffractograms, showing that the spherical and cubic nanocrystals have the same composition.

\begin{figure}[ht]
  \centering
  \includegraphics*[width=0.75\textwidth]{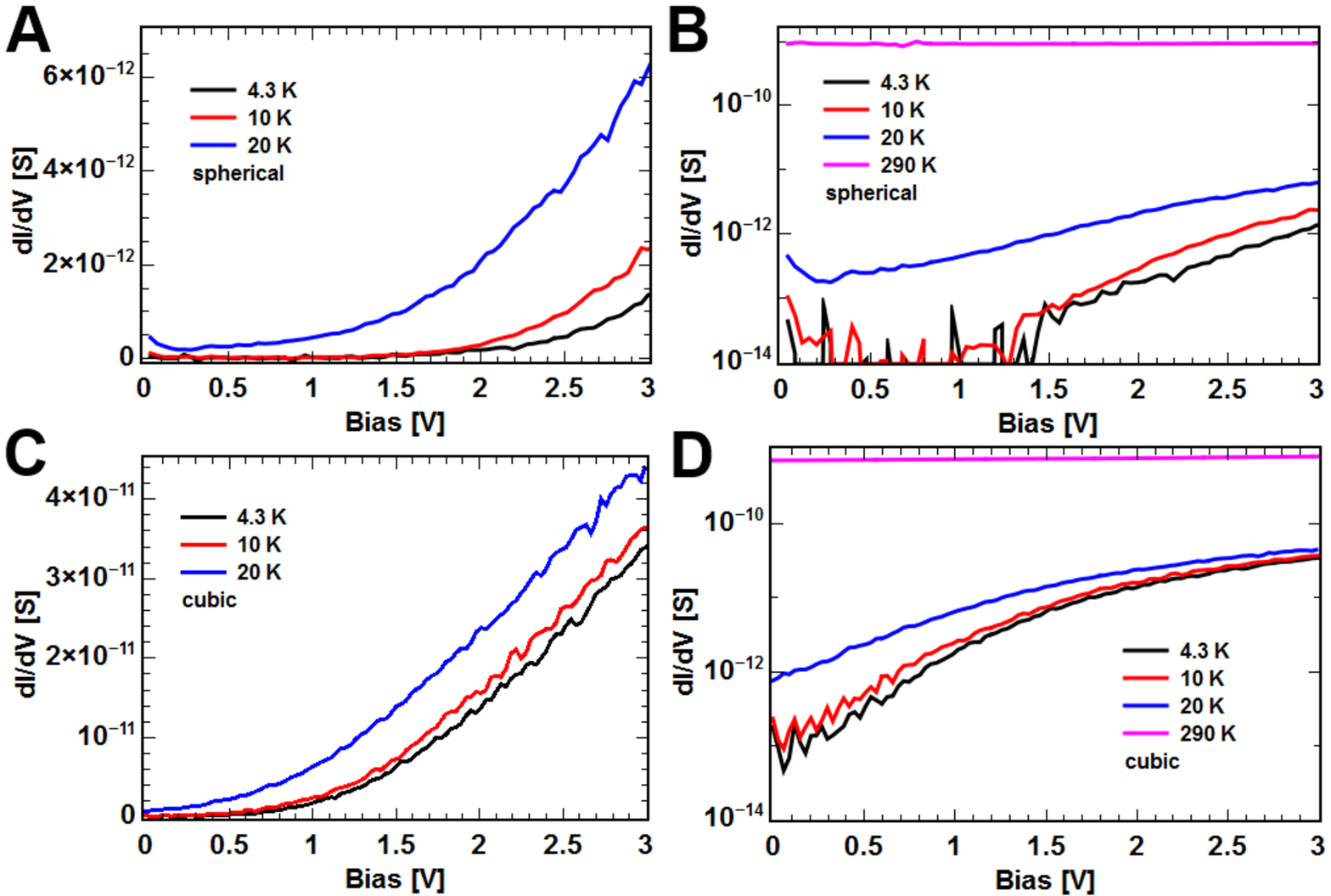}
  \caption{\textit{\textbf{2-terminal measurements.} (A) Conductivity dI$_{D}$/dV$_{DS}$ of a film using spherical nanoparticles. At low temperatures the film is in Coulomb blockade which can better be seen in (B) which shows the same data on a log-scale. (C) shows the data for a film using cubic nanoparticles. (D) shows the data on a log-scale. Both films are covered with 77 nm of Al$_{2}$O$_{3}$.}}
\end{figure}

Classically, spin- and dip-coating \cite{JING2009} are used to prepare films of nanocrystals. Those methods are usually not suitable for high quality monolayers of nanocrystals. Thus, we developed an approach to prepare 2D structures based on the Langmuir-Blodgett method \cite{ALEKSANDROVIC2008,CAI2010,GRESHNYKH2008}. The high quality of the monolayers can be seen in Figure S1 (Supporting Information). In the Langmuir-Blodgett trough the films reach a size of 5 cm $\times$ 7 cm. The monolayer was then deposited onto the surface of a Si/SiO$_{2}$-wafer (with 300 nm thermal oxide) with dimensions of 1 cm $\times$ 1 cm. The substrate was patterned by electron-beam lithography defining gold electrodes with a width of 1 $\mu$m (corresponding to ca. 110 particles) and an inter-electrode distance of 1.2 $\mu$m (corresponding to ca. 130 particles). On top of this structure, we deposited the nanocrystal film by the Langmuir-Blodgett method. Subsequently, we deposited different thicknesses (55 nm, 77 nm, and 150 nm) of aluminum oxide layer by ALD. On top of the aluminum oxide, we performed an additional electron-beam lithography step in order to define a top-gate electrode precisely between the lateral contacts. The whole ensemble works as three-terminal device, a field-effect transistor. The top-gate, in contrast to a back-gate, allows for a more precise electrostatic control of the nanocrystal film. The device concept is shown in Figure 1A-C. The real device was imaged by scanning electron microscopy and is displayed in Figure 1D-E. The lateral electrodes in Figure 1E are somewhat fuzzy since they are underneath the aluminum oxide layer, which is semi-transparent to the electron-beam. In Figure 1F a cross-section is displayed. It nicely shows the assembly of silicon oxide on highly-doped silicon, the important monolayer of metal nanoparticles, the ALD film, and finally the top electrode.

\begin{figure}[!ht]
  \centering
  \includegraphics*[width=0.75\textwidth]{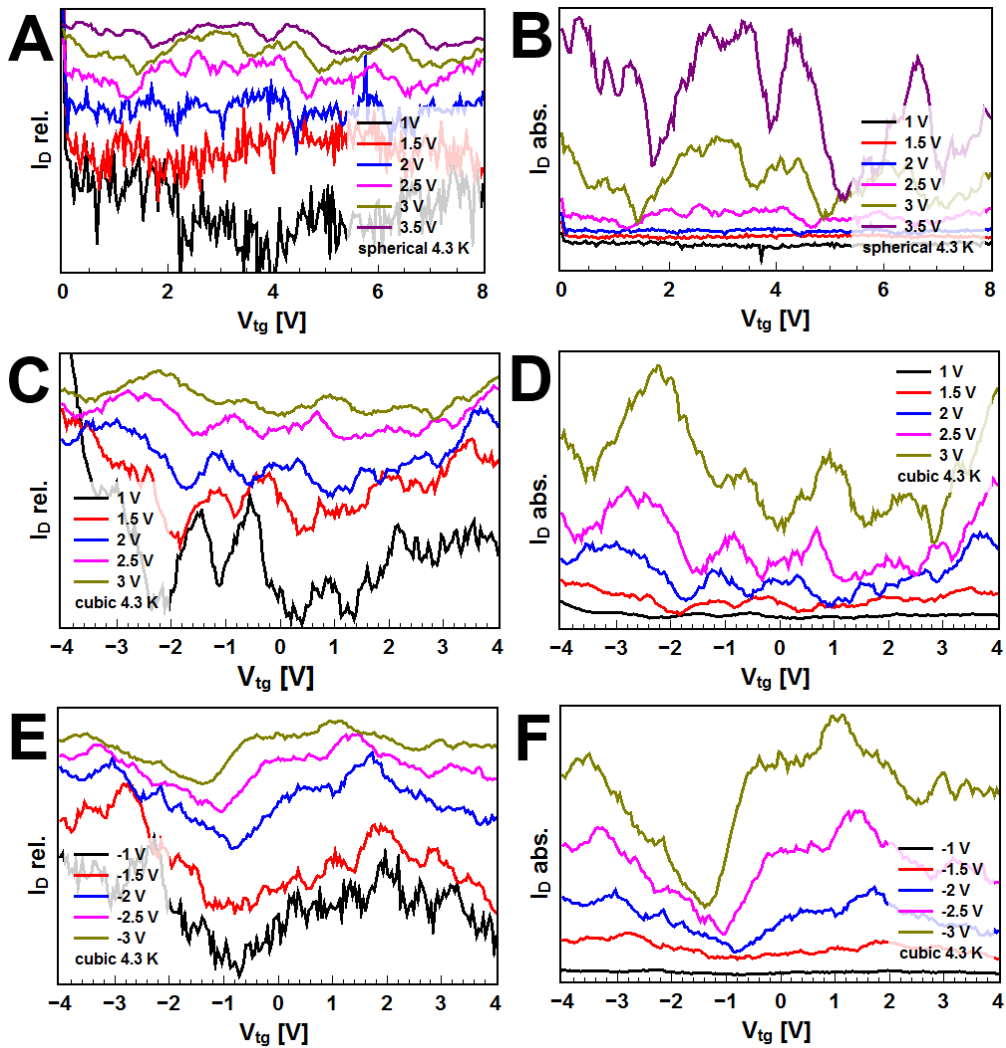}
  \caption{\textit{\textbf{3-terminal measurements of the lateral current as a function of top-gate voltage.} Lateral current I$_{D}$ as a function of the applied top-gate voltage V$_{tg}$. For spherical nanoparticles (smaller size, larger Coulomb charging energy) the oscillations start at a bias V$_{DS}$ of about 2.0 V at 4.3 K (A). For the cubic nanoparticles Coulomb oscillations are already visible at a bias of 1.0 V (C, E). On the left side the relative changes I$_{D}$/I$_{D,0}$ (A, C, E) and on the right side the absolute changes I$_{D}$-I$_{D,0}$ (B, D, F) are shown. In the latter the changes are clear to see not because the relative change is larger but the absolute value of the current is larger at larger biases. The relative current change is larger for smaller value of the bias voltage V$_{DS}$. All measured films are covered with 77 nm Al$_{2}$O$_{3}$. For clarity the curves are shifted on the y-scale. Thus, we do not show scales here but in Figure 5.}}
\end{figure}


\begin{figure}[!ht]
  \centering
  \includegraphics*[width=0.5\textwidth]{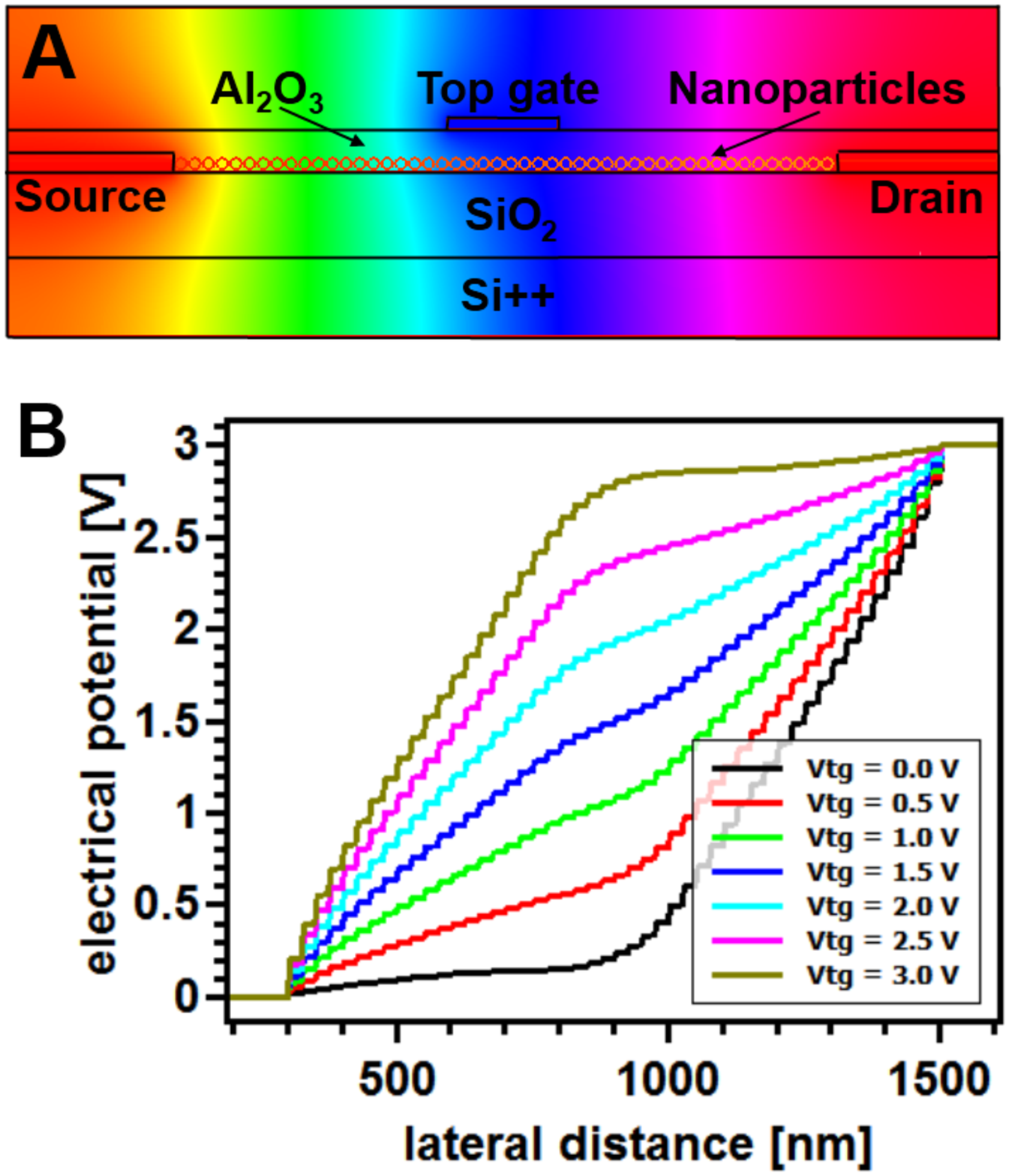}
  \caption{\textit{\textbf{Finite-element method simulations of the electrical potential of the metal nanocrystal transistor.} (A) Cross-section through the device at 1.0 V drain voltage and 0.5 V gate voltage. Red corresponds to a potential of 0.0 V and pink to 1.0 V. (B) Potential along the path of conduction from the left to the right electrode. The left electrode (source) is at 0.0 V and the right one (drain) at 1.0 V. The top-gate voltage is tuned stepwise from 0.0 V to 1.0 V. The back-gate (Si++) is left floating. In the spherical particles the potential is constant leading to stepped functions.}}
\end{figure}

In order to characterize the properties of the constructed devices several electrical measurements have been performed. First, the devices have been evaluated by 2-terminal measurements. The conductance curves dI$_{D}$/dV$_{DS}$ vs. V$_{DS}$ in Figure 2 show clearly that for spherical nanoparticles and temperatures below 10 K the film is completely in Coulomb blockade at biases below about 1.5 V (the corresponding I$_{DS}$-V$_{DS}$ curves are shown in Figure S3 in the Supporting Information). At higher temperatures the Coulomb blockade bias is reduced and at temperatures over 20 K it is completely lifted. Using the larger cubic nanoparticles the Coulomb blockade bias lies roughly at 0.5 V at 4.3 K. This is due to the lower activation energy for tunneling across the gaps between the cubic nanoparticles, in accordance with our earlier publication \cite{GRESHNYKH2008}.

\begin{figure}[!ht]
  \centering
  \includegraphics*[width=0.75\textwidth]{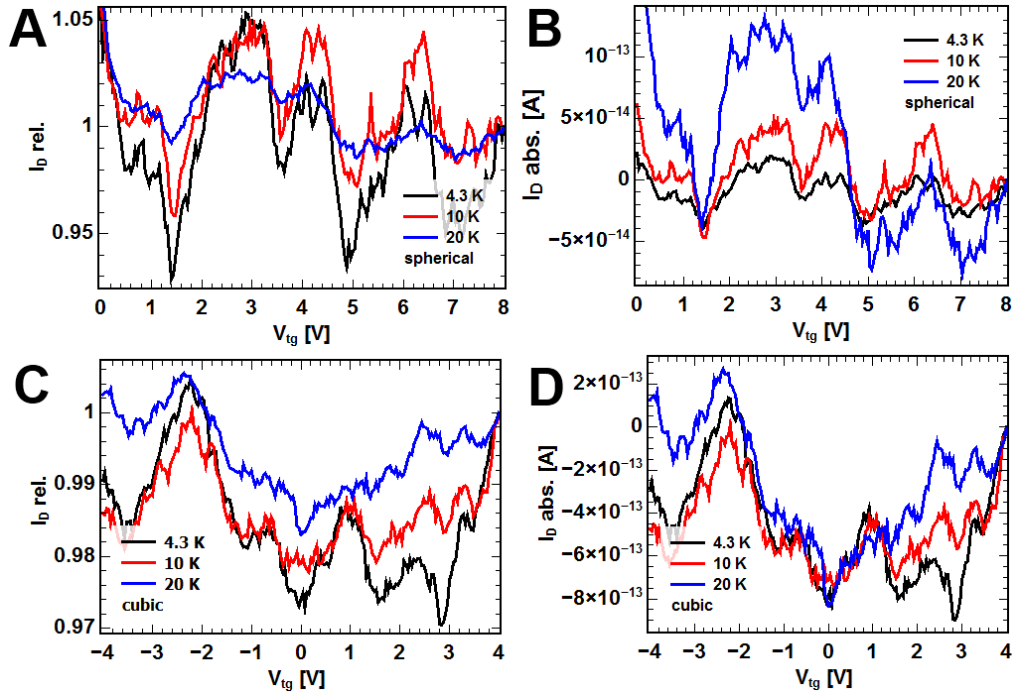}
  \caption{\textit{\textbf{3-terminal measurements of the lateral current as a function of temperature.} Lateral current I$_{D}$ as a function of temperature at a bias of +3.0 V with spherical (A, B) and cubic (C, D) nanocrystal films (both covered with 77 nm of Al$_{2}$O$_{3}$). (A, C) relative change I$_{D}$/I$_{D,0}$; (B, D) absolute change I$_{D}$-I$_{D,0}$. The relative changes become weaker with increasing temperature and vanish above 20 K. This is more pronounced for the smaller particles. The absolute changes become more pronounced with increasing temperature. Both aspects are due to an overall increase of the current with temperature.}}
\end{figure}

The corresponding current-voltage curves can be fitted with the threshold voltage model from Middleton et al. \cite{MIDDLETON1993}. We fit the current-voltage curves of spherical and cubic nanoparticles measured at 4.3 K and determine the $\zeta$ values and the threshold voltages (see Figure S4 in the Supporting Information). For the spherical nanoparticles we obtain a $\zeta$ value of about 3.0 and a threshold voltage of 1.6 V. The cubic nanoparticles show a threshold voltage at 0.6 V and we obtain a $\zeta$ value of 2.6. The different $\zeta$ values of the two different particle shapes are probably caused by the different packing of the film. The deviation in the threshold voltages is caused by the different Coulomb energies of the nanoparticles \cite{GRESHNYKH2008}. The $\zeta$ values are much higher than theory predicts \cite{MIDDLETON1993} but in agreement with the experimental values from other groups \cite{ANCONA2001,BEECHER2005,TAN2009}. For metallic 2D nanocrystal arrays $\zeta$ values of 2 to 2.5 were determined, depending on the size distribution of the nanoparticles. For multilayer nanocrystal films $\zeta$ values of about 2.6 to 3 were determined \cite{BLACK2000,ANCONA2001,BEECHER2005,TAN2009}. Gold nanocrystal arrays with strong topological inhomogeneities show $\zeta$ values of about 4 \cite{BLUNT2007}.

Calculations and experimental results suggest that the charging energy only depends on the particles' self-capacitance (C$_{\text{self}}$) and is almost independent of the inter-particle distance \cite{GRESHNYKH2008}. Thus, the charging energy is independent of the geometric capacitance, as well. This was demonstrated by calculations using the finite-element method. The geometrical capacitance starts to be predominant over the self-capacitance when the particles' diameter exceeds one micrometer. Therefore, the charging energy of smaller particles only depends on the dielectric constant ($\epsilon_{r}$) of the interparticle medium and on the particles' radius. The dielectric constant of the interparticle medium was calculated by using the activation energy:

\begin{equation*}
E_{a} = \frac{e^{2}}{2 C_{\text{self}}} = \frac{e^{2}}{8 \pi \epsilon_{o} \epsilon _{r} R} \hspace{6 mm} \Leftrightarrow \hspace{6 mm} \epsilon _{r} = \frac{e^{2}}{8 \pi \epsilon_{o} R E_{a}}
\end{equation*}

The activation energy is deducted from the slope of a corresponding Arrhenius plot. For spherical nanocrystal of 7.6 nm in diameter an activation energy of 20 meV and a dielectric constant of 9.5 was calculated. The self-capacitance of the spherical particle is thus 4.0 aF. These values are on the same order of magnitude as the values of earlier publications \cite{GRESHNYKH2008,BEECHER2005}.

Further, we measured the lateral current I$_{D}$ through the nanocrystal films with a layer of 77 nm Al$_{2}$O$_{3}$ as a function of the top-gate electrode V$_{tg}$ from -4 V to +4 V (or from 0 V to +8 V) while stepping the bias V$_{DS}$ from -3 V to +3 V in 0.5 V steps. Figure 3 shows on the left side (A, C, E) the relative change of the current and on the right side (B, D, F) the absolute change. The relative change of the current is defined by I$_{D}$ rel. = I$_{D}$/I$_{D,0}$, where I$_{D,0}$ is the last values of the current I$_{D}$ in the measured range. The absolute change of current is defined by I$_{D}$ abs. = I$_{D}$-I$_{D,0}$. Those current variations are defined in order to increase the visibility of the effects. (The raw data for the source, drain, and top-gate current are exemplarily shown for a bias of -3 V in Figure S5 in the Supporting Information.) In the relative current plots the changes are more clearly observable for small biases V$_{DS}$, and in the absolute current plots the changes for larger biases are more obvious. At a temperature of 4.3 K, the film using spherical nanoparticles shows Coulomb oscillations at a bias of 2.0 V and above. With cubic nanocrystals the Coulomb oscillations are already visible at a lower bias of 1.0 V. These values are in good agreement with the values of the threshold voltage and the Coulomb blockade values from the current-voltage curves (Fig. 2). The peaks of the Coulomb oscillations are non-periodic, in accordance with the sequential tunneling of the electrons through multiple particles and pathways \cite{JOUNG2011}. In relative terms, the Coulomb oscillations can be seen most clearly just above the threshold voltage, since at these conditions the electrical field is sufficient to overcome the Coulomb blockade through the whole film for the first time and the charge transport most probably takes place along a single path or a small number of branches. If the bias is increased, the current flows through multiple pathways, which branch out and reconnect \cite{MIDDLETON1993}, and the Coulomb oscillations become weaker. In absolute values, the oscillations are more pronounced for higher biases due to an overall higher current.

\begin{figure}[ht]
  \centering
  \includegraphics*[width=0.75\textwidth]{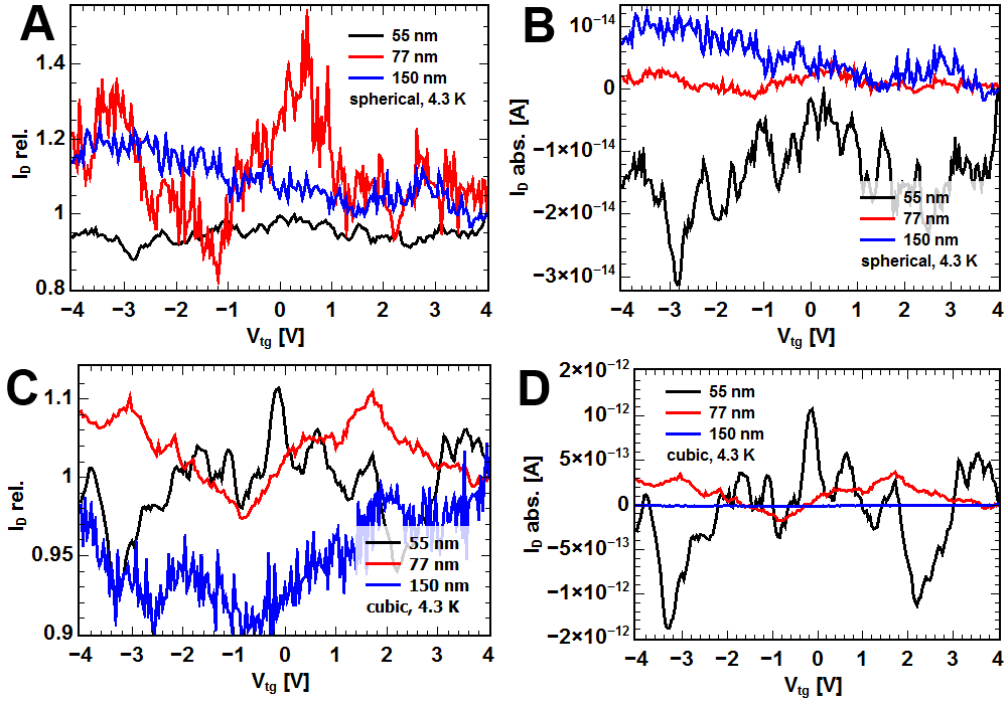}
  \caption{\textit{\textbf{3-terminal measurements of the lateral current as a function of gate oxide thickness.} Lateral current I$_{D}$ as a function of the Al$_{2}$O$_{3}$ layer thickness at a bias of -2 V with spherical (A, B) and (C, D) cubic nanoparticles. With 55 nm and 77 nm of Al$_{2}$O$_{3}$ the oscillations are clearly visible. With 150 nm of Al$_{2}$O$_{3}$ the oscillations become noisy in the relative change and are not to see in the absolute change.}}
\end{figure}

The oscillations in the I$_{D}$-V$_{tg}$ curves can be understood by means of simulations of the electrostatic potential using the finite-element method. Figure 4 shows in (A) the distribution of the electric potential at +1.0 V bias and +0.5 V top-gate voltage and in (B) the potential along the conductive path is shown for different top-gate voltages, showing that the local electric potential and the electric field can nicely be tuned by a local top-gate. We attribute the oscillations in the I$_{D}$-V$_{tg}$ curves to a change in the local electric potential. 

The features in the I$_{D}$-V$_{tg}$ curves shift to more positive gate voltages with increasing bias V$_{DS}$ and to more negative gate voltages with decreasing bias, which is in accordance with the picture of sequential tunnelling through multiple particles. At the half value of the bias V$_{tg}$ = V$_{DS}$/2 the currents always show a local minimum. This is because at this top-gate voltage the maximum occurring electric field strength has a minimum.

For the temperature measurements we also swept the top-gate electrode voltage V$_{tg}$ from -4 V to +4 V (or from 0 V to +8 V). Exemplarily, we show the curves at a bias of +3 V (Fig. 5). With increasing measurement temperature the Coulomb oscillations become weaker. In the relative value plots, the oscillations vanish completely above 20 K. At elevated temperature, conductance through the nanocrystal films is no longer determined by the Coulomb blockade. In absolute terms, the oscillations become more pronounced with increasing temperature due to a higher current level, but also vanish above 20 K. The maximum temperature to observe Coulomb oscillations is 20 K. This is in accordance with the lift of the Coulomb blockade regime observed in the I$_{D}$-V$_{DS}$ curves (Fig. 2). Figure 5A also shows that the modulation of the lateral current by the top-gate is on the order of $\pm$5 \%.

We also characterized samples with different Al$_{2}$O$_{3}$ layer thicknesses, namely 55 nm, 77 nm, and 150 nm. This layer should be as thin as possible to ensure the strongest effect of the top-gate, but it has to be thick enough to prevent leakage current to the top-gate. The optimum layer thickness has to be determined experimentally. For the measurement we sweep again the top-gate voltage from -4 V to +4 V and set the bias from $\pm$1 V to $\pm$3 V. Exemplarily, we show the curves at -2 V because the curves at this voltage are least noisy. With increasing layer thickness the Coulomb oscillations become weaker and the oscillation frequencies decrease. This can be seen most clearly in the absolute value plots as shown in Figure 6. With 55 nm and 77 nm of Al$_{2}$O$_{3}$ thickness, Coulomb oscillations are clearly visible in the current plots. The Coulomb oscillations of the film with 150 nm of Al$_{2}$O$_{3}$ becomes noisy in the relative value plot and are not observable at all in at the absolute value plot. With 55 nm of Al$_{2}$O$_{3}$, the oscillations are most precisely defined and the oscillation frequencies is the highest, since the gate electrode is closest to the film. Thus, the local potential at the particles is influenced most directly. In some measurements with 55 nm of oxide a small leakage current to the top-gate was detectable, but for larger oxide layers the leakage currents are smaller than the noise level. The thinner the oxide layer is, the more pronounced are the features in the plots, but with decreasing oxide layer the gate leakage current increased. The variation of the oxide layer thickness suggests an ideal Al$_{2}$O$_{3}$ thickness of 77 nm. At this thickness the oscillations are clearly visible and the leakage current is within the noise level of the setup of 10$^{-14}$ A. With 150 nm of Al$_{2}$O$_{3}$ the distance to the film is too large, so that the effect of the gate is barely measurable. The effect of top-gate distance could also be observed in finite-element simulations. They show that the highest field along the path through the nanoparticles under the top-gate decreases sharply with increasing distance (Figure S6).


To conclude, the electrical measurements conducted in this work show that a Coulomb blockade based transistor using a two-dimensional colloidal metallic nanocrystal film is possible. For that, we used well-separated, well-organized nanoparticles in order to maintain the Coulomb blockade features of individual nanoparticles. Anyhow, the fixed inter-particle distance of about 1 nm allows for an overlap of the wave function to enable tunneling. Due to the film height of only one monolayer of nanocrystals screening is limited and Coulomb oscillations were observed in the current vs. top-gate voltage plots. With increasing measurement temperature, the oscillations weaken and disappear above 20 K. It was also found that with increasing distance between a top-gate electrode and the nanocrystal film, the effect of the top-gate voltage decreased. A change of bias led to a shift of the oscillation peaks. This was attributed to a change of the local electric potential. The prepared nanocrystal film devices using different batches of synthesized nanoparticles, with different particle shapes and sizes behave very reproducible and according to theory. They are stable over time due to the protective oxide layer which at the same time is the top-gate dielectric. The processing of the metal nanocrystal transistor is comparatively simple, inexpensive, and CMOS compatible, whereat the particular transfer characteristics can lead to applications exhibiting new functionality. For the future, it should be possible to increase the temperature of operation by the use of even smaller nanoparticles.

\section*{Acknowledgments}

The authors acknowledge the German Research Foundation DFG and the European Research Council (Seventh Framework Program FP7) for funding.




\begin{thebibliography}{00}

\bibitem{FULTON1987} T. A. Fulton and G. J. Dolan, Phys. Rev. Lett. \textbf{59}, 109 (1987).
\bibitem{ANDRES1997} R. P. Andres, T. Bein, M. Dorogi, S. Feng, J. I. Henderson, C. P. Kubiak, W. Mahoney, R. G. Osifchin, and R. Reifenberger, Science \textbf{272}, 1323 (1996).
\bibitem{KLEIN1996} D. L. Klein, P. L. McEuen, J. E. Bowen Katari, R. Roth, and A. P. Alivisatos, Appl. Phys. Lett. \textbf{68}, 2574 (1996).
\bibitem{BEZRYADIN1997} A. Bezryadin, C. Dekker, and G. Schmid, Appl. Phys. Lett. \textbf{71}, 1273 (1997).
\bibitem{WOLF2010} C. R. Wolf, K. Thonke, and R. Sauer, Appl. Phys. Lett. \textbf{96}, 142108 (2010).
\bibitem{SATO1997} T. Sato, H. Ahmed, D. Brown, and B. F. G. Johnson, J. of Appl. Phys. \textbf{82}, 696 (1997).
\bibitem{BOLOTIN2004} K. I. Bolotin, F. Kuemmeth, A. N. Pasupathy, and D. C. Ralph, Appl. Phys. Lett. \textbf{84}, 3154 (2004).
\bibitem{LIKHAREV1999} K. K. Likharev, Proceedings of the IEEE \textbf{87}, 606 (1999).
\bibitem{TALAPIN2005} D. V. Talapin and C. B. Murray, Science \textbf{310}, 86 (2005).
\bibitem{LIAO2006} J. Liao, L. Bernard, M. Langer, C. Schonenberger, and M. Calame, Adv. Mater. \textbf{18} 2444 (2006).
\bibitem{MURRAY2007} J. J. Urban, D. V. Talapin, E. V. Shevchenko, C. R. Kagan, and C. B. Murray, Nat. Mater. \textbf{6}, 115 (2007).
\bibitem{TALAPIN2009}	D. V. Talapin, J. G. Lee, M. V. Kovalenko, and E. V. Shevchenko, Chem. Rev. \textbf{110}, 389 (2009).
\bibitem{BLACK2000} C. T. Black, C. B. Murray, R. L. Sandstrom, and S. Sun, Science \textbf{290}, 1131 (2000).
\bibitem{STAVEREN1991} M. P. J. van Staveren, H. B. Brom, and L. J. de Jongh, Phys. Rep. \textbf{208}, 1 (1991).
\bibitem{PENG1999} D. L. Peng, K. Sumiyama, T. J. Konno, T. Hihara, and S. Yamamuro, Phys. Rev. B \textbf{60}, 2093 (1999).
\bibitem{BEVERLY2002}	K. C. Beverly, J. L. Sample, J. F. Sampaio, F. Remacle, J. R. Heath, and R. D. Levine, Proc. Natl. Acad. Sci. USA \textbf{99}, 6456 (2002).
\bibitem{BEECHER2004}	P. Beecher, A. J. Quinn, E. V. Shevchenko, H. Weller, and G. Redmond, J. Phys. Chem. B \textbf{108}, 9564 (2004).
\bibitem{ANDRES2006} R. P. Andres, J. D. Bielefeld, J. I. Henderson, D. B. Janes, V. R. Kolagunta, C. P. Kubiak, W. J. Mahoney, R. G. Osifchin, Science \textbf{273}, 1690 (1996).
\bibitem{MARKIVICH1999}	G. Markovich, C. P. Collier, S. E. Henrichs, F. Remacle, R. D. Levine, and J. R. Heath, Acc. Chem. Res. \textbf{32}, 415 (1999).
\bibitem{SIMON1998} U. Simon, Adv. Mater. \textbf{10}, 1487 (1998).
\bibitem{QUINN2005}	A. J. Quinn, P. Beecher, D. Iacopino, L. Floyd, G. De Marzi, E. V. Shevchenko, H. Weller, and G. Redmond, Small \textbf{1}, 613 (2005). 
\bibitem{GIAEVER1968}	I. Giaever and H. R. Zeller, Phys. Rev. Lett. \textbf{20}, 1504 (1968).
\bibitem{GORTER1951} C. J. Gorter, Physica \textbf{17}, 777 (1951).
\bibitem{LAMBE1969} J. Lambe and R. C. Jaklevic, Phys. Rev. Lett. \textbf{22}, 1371 (1969).
\bibitem{ABELES1975} B. Abeles, P. Sheng, M. D. Coutts, and Y. Arie, Adv. Phys. \textbf{24}, 407 (1975).
\bibitem{INGOLD1992} G. L. Ingold and Y. V. Nazarov, NATO ASI Series B \textbf{294}, 21 (1992).
\bibitem{NEUGEBAUER1962} C. A. Neugebauer and M. B. Webb, J. Appl. Phys. \textbf{33}, 74 (1962).
\bibitem{MIDDLETON1993}	A. A. Middleton and N. S. Wingreen, Phys. Rev. Lett. \textbf{71}, 3198 (1993).
\bibitem{SUVAKOV2010} M. Suvakov and B. Tadic, B., J. Phys. Cond. Mat. \textbf{22}, 163201 (2010).
\bibitem{PARTHASARATHY2001} R. Parthasarathy, X. M. Lin, and H. M. Jaeger, Phys. Rev. Lett. \textbf{87}, 186807 (2001).
\bibitem{PARTHASARATHY2004} R. Parthasarathy, X. M. Lin, K. Elteto, T. F. Rosenbaum, and H. M. Jaeger, Phys. Rev. Lett. \textbf{92}, 076801 (2004).
\bibitem{TAO2008}	A. R. Tao, J. Huang, and P. Yang, Acc. Chem. Res. \textbf{41}, 1662 (2008).
\bibitem{WELLER2003}	H. Weller, Phil. Trans. R. Soc. A \textbf{361}, 229 (2003).
\bibitem{AHRENSTORF2007} K. Ahrenstorf, O. Albrecht, H. Heller, A. Kornowski, D. Gorlitz, and H. Weller, Small \textbf{3}, 271 (2007).
\bibitem{PARK2007} J. Park, J. Joo, S. G. Kwon, Y. Jang, and T. Hyeon, Angew. Chem. Int. Ed. \textbf{46}, 4630 (2007).
\bibitem{SHEVCHENKO2003} E. V. Shevchenko, D. V. Talapin, H. Schnablegger, A. Kornowski, O. Festin, P. Svedlindh, M. Haase, and H. Weller, J. Am. Chem. Soc. \textbf{125}, 9090 (2003).
\bibitem{MALYNYCH2002} S. Malynych, I. Luzinov, and G. Chumanov, J. Phys. Chem. B \textbf{106}, 1280 (2002).
\bibitem{ALEKSANDROVIC2008} V. Aleksandrovic, D. Greshnykh, I. Randjelovic, A. Fromsdorf, A. Kornowski, S. V. Roth, C. Klinke, and H. Weller, ACS Nano \textbf{2}, 1123 (2008).
\bibitem{ARAI2005} T. Arai, S. Saito, H. Fukuda, and T. Onai, Jpn. J. Appl. Phys. \textbf{44}, 5667 (2005).
\bibitem{HOSOKI2008} K. Hosoki, T. Tayagaki, S. Yamamoto, K. Matsuda, and Y. Kanemitsu, Phys. Rev. Lett. \textbf{100}, 207404 (2008).
\bibitem{CAI2010} Y. Cai, D. Wolfkuhler, A. Myalitsin, J. Perlich, A. Meyer, and C. Klinke, ACS Nano \textbf{5}, 67 (2010).
\bibitem{GRESHNYKH2008} D. Greshnykh, A. Fromsdorf, H. Weller, and C. Klinke, Nano Lett. \textbf{9}, 473 (2008).
\bibitem{SHEVCHENKO2002} E. V. Shevchenko, D. V. Talapin, A. L. Rogach, A. Kornowski, M. Haase, and H. Weller, J. Am. Chem. Soc. \textbf{124}, 11480 (2002).
\bibitem{JING2009} P. Jing, J. Zheng, Q. Zeng, Y. Zhang, X. Liu, X. Liu, X. Kong, and J. Zhao, J. Appl. Phys. \textbf{105}, 044313 (2009).
\bibitem{GEORGE2009} S. M. George, Chem. Rev. \textbf{110}, 111 (2009).
\bibitem{PUURUNEN2005} R. L. Puurunen, J. Appl. Phys. \textbf{97}, 121301 (2005).
\bibitem{ANCONA2001} M. G. Ancona, W. Kruppa, R. W. Rendell, A. W. Snow, D. Park, and J. B. Boos, Phys. Rev. B \textbf{64}, 033408 (2001).
\bibitem{BEECHER2005}	P. Beecher, E. V. Shevchenko, H. Weller, A. J. Quinn, and G. Redmond, Adv. Mater. \textbf{17}, 1080 (2005).
\bibitem{TAN2009} R. P. Tan, J. Carrey, C. Desvaux, L. M. Lacroix, P. Renaud, B. Chaudret, and M. Respaud, Phys. Rev. B \textbf{79}, 174428 (2009).
\bibitem{BLUNT2007} M. O. Blunt, M. Suvakov, F. Pulizzi, C. P. Martin, E. Pauliac-Vaujour, A. Stannard, A. W. Rushforth, B. Tadic, and P. Moriarty, Nano Lett. \textbf{7}, 855 (2007).
\bibitem{JOUNG2011} D. Joung, L Zhai, and S. I. Khondaker, Phys. Rev. B \textbf{83}, 115323 (2011).

\end{thebibliography}
\end{document}